\documentclass[onecolumn,showpacs,preprintnumbers,amsmath,amssymb]{revtex4}

\usepackage{amsmath,amssymb}
\usepackage{natbib}
\usepackage{url}
\usepackage{graphicx}
\usepackage{psfrag}
\usepackage{color}
\usepackage{epsfig}

\begin{document}

\title{One dimensional description of the gravitational perturbation in a Kerr background}
\author{Dar\'{\i}o N\'u\~nez$^{1}$, Juan Carlos Degollado$^{1}$ and Carlos Palenzuela$^{2,3}$}

\date{\today}

\label{firstpage}
\affiliation{$^{1}$Instituto de Ciencias Nucleares, Universidad
Nacional Aut\'onoma de M\'exico, Apdo. 70-543, CU, 04510 M\'exico,
D. F., M\'exico.\\
$^{2}$Canadian Institute for Theoretical Astrophysics, University of Toronto,
M5S 3H8 Toronto, Ontario, Canada. \\
$^{3}$Max-Planck-Institut f\"ur Gravitationsphysik, Albert Einstein Institut,
14476 Golm, Germany} 
\email{nunez@nucleares.unam.mx, jcdegollado@nucleares.unam.mx,
carpa@aei.mpg.de}

\begin{abstract}
We find a  way to write the perturbation equation in a Kerr background
as a coupled system of one dimensional equations for the different modes 
in the time domain. Numerical simulations show that the dominant mode in the
gravitational response is the one corresponding to the mode of the initial 
perturbation, allowing us to conjecture that the coupling among the modes
has a weak influence in our one dimensional system of equations.
We conclude that by neglecting the coupling terms it can be obtained
a one dimensional harmonic equation which indeed describes
with good accuracy the gravitational response from the Kerr black hole.
This result may help to understand the structure of test fields in a Kerr
background and even to generate accurate waveforms for various cases in
an efficient manner.
\end{abstract}



\maketitle

\section{Introduction}

There is an intrinsic interest on understanding the structure of black holes
and the behavior of the fields close to them, where the gravity effects are strong.
One natural way to describe such a system is through the linearization
of the Einstein equations around a fixed background solution.
From the theoretical point of view, linear perturbations allow us to study
the stability of the spacetime, the damping of the perturbations at late times (ie, the
Quasi-Normal-Modes and the tails)
and other phenomena like the super radiance or the extraction
of rotational energy from the black hole. On a more practical side, it can be used as
a tool for computing the gravitational wave emission when the spacetime is perturbed by a
less massive source. In these cases the spacetime is dominated by the black hole and it
can be considered as a fixed background. A very likely occurrence example is the capture
of compact stellar-sized objects 
(ie, a black hole or a star) by the super massive black holes lying at the center of a galaxy.
This type of Extreme Mass Ratio Inspiral (EMRI) events will generate gravitational waves within
the sensitivity band of LISA, the prospected space-based GW detector. In spite of the
recent breakthroughs done by Numerical Relativity in the full 3D evolution
of binary black holes,
this is one of the problems which seems not possible to be solved easily by these methods, due
mainly to the very different scales presents on the problem.

One of the frames to study the perturbations is by means the linearization
of the Einstein Equations through the perturbed Weyl scalar ${\Psi_4}$
\cite{Teu73}. The perturbations of the Schwarzschild solution (i.e., non-rotating BH)
are well known since the seventies \cite{Price72}. In this case,
the solution of the angular part is simply given by the spin weighted
spherical harmonics, while the time-radial part describes
the evolution of multipoles which evolve independently.

Perturbations around Kerr spacetime (i.e., rotating BH) are much more complicated,
since the spacetime is not spherically symmetric and the fields can not be
decomposed into independently evolving multipoles in the time domain.
Already Teukolsky \cite{Teu73}, with a null tetrad of the Kinnersley type, arrived to a
separable master perturbation equation for the Kerr black hole in Boyer-Lindquist
coordinates. However, the angular part was described by the so called spheroidal
spin weighted harmonics, for which there is no known analytical expression.
Moreover, the multipole decomposition could only be performed in the
frequency domain, but with separation constants which were functions
of frequency, implying a non-trivial (and unknown) multipole interaction in 
the time domain. A expansion on the spin parameter was used in \cite{Gleiser08}
to separate the modes and study the effects of the mode coupling
on the tails, the very late behaviour of the perturbations after
the ring down. However, their results are by construction restricted to
small deviations from Schwarzschild spacetime (see also \cite{Burko09}).

In this work, we describe the Kerr black hole in horizon penetrating
coordinates and define a new tetrad such that the main angular operator, in
the evolution equation for the Newman-Penrose scalar ${\Psi_4}$,
has as eigenfunctions the usual spin weighted spherical harmonics.
This allows us to write the ${\Psi_4}$ in a base of these
harmonics, without the need to define the spheroidal ones. The action of
the main angular operator on ${\Psi_4}$ can be expressed in terms of
well known eigenvalues. This procedure leads to an operator with only radial and temporal 
derivatives, but still with angular coefficients. By using the
normalization properties of the spherical harmonics, it is possible to arrive
to a purely radial-temporal operator for a combination of modes, reducing in this 
way the perturbation problem in Kerr to a coupled system of one dimensional
equations for the different modes. A similar procedure can be performed
with the Maxwell and the Klein-Gordon equations.

Furthermore, our procedure allows to get a clearer description of the
modes composing the gravitational wave in terms of the initial data. We anticipate
that the dominant mode in the waveform is the one corresponding to the mode of the
initial data, implying that the coupling terms are weak with respect to the
main harmonic operator. This result further reduces the problem of generating wave
forms in a perturbed Kerr spacetime, since up to some level of accuracy
is enough to simply solve the equation for the corresponding mode.

The work is organized as follows: in section \ref{sec:pert_eq}, we present a derivation of 
the perturbation equation for the perturbed $\Psi_4$ Weyl scalar for a Kerr black hole described in
Kerr-Schild penetrating coordinates, and explicitly derived the evolution equation for
$\Phi=r\,{\Psi_4}$. In section \ref{sec:decomposition} the function $\Phi$
is decomposed in terms of spin weighted spherical harmonics. After some manipulations,
the perturbation equation can be separated in radial-temporal and angular parts, resulting in 
a coupled system of radial-temporal equations. In section \ref{sec:firstorder} it is
shown a first order reduction of the coupled system of equations. It is also explained
the standard approach, which consists on using a different base of angular functions
and solve the Teukolsky equation in 2D. In section \ref{sec:evolution} we describe the numerical 
code used to evolve the two previous systems and present few numerical 
examples of the gravitational response of the black hole due to a gravitational pertubation.
Finally 
in section \ref{sec:conclusion} we present a discussion of the results obtained.

\section{Perturbation equation in Kerr spacetime}\label{sec:pert_eq}

The derivation of an evolution equation for the perturbed Weyl scalar requires first
the choice of a convenient null tetrad ${Z_a}^\mu$. Following a previous work \cite{JC09}, we 
define a null tetrad with two real future-directed null vectors ${Z_0}^\mu=l^\mu$ and
${Z_1}^\mu=k^\mu$, where $l^\mu$ is pointing outward and $k^\mu$ pointing inward in an
hypersurface of constant time. The other two null vectors lay in the plane perpendicular to the 
light cone and will be denoted by a complex vector $m^\mu$, such that ${Z_2}^\mu=m^\mu,
{Z_3}^\mu={m^*}^\mu$. These null vectors must satisfy the following equations:
\begin{equation}
{Z_a}^\mu\,{Z_b}_\mu=\eta_{ab}, \hspace{1cm} g_{\mu\nu}=2\eta^{ab}\,Z_{a\,(\mu}\,Z_{b\,\nu)},\label{eqs:tetrad}
\end{equation}
being $\eta_{ab}$ a matrix where all the coefficients are zero except
$\eta_{12}=\eta_{21}=-\eta_{34}=-\eta_{43}=\eta$. The constant $\eta$ is equal to one if the
spacetime 
has signature $\{+,-,-,-\}$, and equal to minus one for opposite signature $\{-,+,+,+\}$.

Following the procedure and definitions described in \cite{JC09,MoNun01}, we obtain the
following evolution equation for the perturbed Weyl scalar ${\Psi_4}^{(1)}$, for type D spaces in vacuum backgrounds, 
\begin{eqnarray}
&&\left(\left({\bf \Delta} + \eta\, \left(4\,\mu + \mu^* + 3 \gamma -\gamma^* \right)\right)\,
\left({\bf D} - \eta\,\left(\rho -4\,\epsilon \right) \right) - \left({\bf \delta^*} + \,
\left(3 \alpha + \beta^* + 4\,\pi - \tau^* \right)\right)\,\left( {\bf \delta} +
\eta\,\left(4\,\beta-\tau \right) \right) \right.  \nonumber \\ - &&\left. 3\,\eta_{12}\,\Psi_2
\right)\,{\Psi_4}^{(1)}
=\eta\,4\,\pi\,T \label{eq:PertPsi4f}
\end{eqnarray}
where we have used geometric units. This evolution equation for the perturbed Weyl scalar was
originally derived by Teukolsky \cite{Teu73}, modulus the coefficient $\eta$ which takes into
account the signature convention. It was successfully used to describe, in the frequency domain, the
gravitational waves generated by a perturbed rotating black hole.


Now we proceed to use the perturbation equation (\ref{eq:PertPsi4f}) in a background
given by the Kerr black  hole described in penetrating coordinates. In these coordinates, the line element is given by:
\begin{eqnarray}
ds^2=&&-\left(1-2\frac{M\,r}{\Sigma}\right)\,dt^2 + 4\,\frac{M\,r}{\Sigma}\,dt\,dr 
-  4\,\frac{M\,r}{\Sigma}\,a\,\sin^2\theta\,dt\,d\varphi +
\left(1+2\frac{M\,r}{\Sigma}\right)\,dr^2 
- 2\,\left(1+2\frac{M\,r}{\Sigma}\right)\,a\,\sin^2\theta\,dr\,d\varphi + \nonumber \\
&& \Sigma\,d\theta^2
+  \left(r^2 + a^2 +2\frac{M\,r}{\Sigma}\,a^2\,\sin^2\theta\right)\,\sin^2\theta\,d\varphi^2 \label{eq:lel_kerr}
\end{eqnarray}
with $M$ is the mass of the black hole, $a$ is its angular momentum per unit mass, and we have 
defined $\Sigma=r^2 + a^2\,\cos^2\theta$. Notice that we have already chosen our signature so that
$\eta=-1$.

The choice of the null tetrad is ambigous, since it is only restricted by the
eqs.~(\ref{eqs:tetrad}) 
with $\eta=-1$. Based on the procedures used
in \cite{JC09, Zeng08a} for the Schwarzschild case, we define the following tetrad:
\begin{equation}
l^\mu=\frac1{2\,\Sigma}\left(r^2+a^2+2\,M\,r,\Delta,0,2\,a\right)~,
\hspace{0.5cm} k^\mu=k_0\,\left(1,-1,0,0\right)~,
\hspace{0.5cm} m^\mu=\frac1{\sqrt{2}\,\left(r-i\,a\,\cos\theta\right)}
\left(i\,a\,\sin\theta,0,1,i\,\csc\theta\right), \label{tet_kerr}
\end{equation}
where we have used the shortcut $\Delta=r^2+a^2-2\,M\,r$. In
the spatial asymptotic region (i.e., $r \to \infty$), the real
vectors of the null tetrad (\ref{tet_kerr}) take the form expected in
a flat spacetime $l^\mu=\frac1{2}\left(1,1,0,0\right),
k^\mu=k_0\,\left(1,-1,0,0\right)$. Regarding the complex null
vector, our choice differs slightly from the one commonly used
in the literature \cite{Teu73}, changing the denominator by its
conjugate complex. This rotation will have a big impact on simplifying
the final perturbation equation.

From the definitions of the spinor coefficients \cite{JC09} it is straighforward
to obtained that the un-perturbed Weyl scalars are $\Psi_0=\Psi_1=\Psi_3=\Psi_4=0$
and $\Psi_2=M\,\mu^3$. The only non-vanishing spinor coefficients in this case are:
\begin{eqnarray}
&\mu=\frac1{r-i\,a\,\cos\theta}; \hspace{0.5cm} 
\rho=\frac{\Delta}{2\,\Sigma}\,\mu; \hspace{0.5cm} 
\epsilon=\rho- \frac{r-M}{2\,\Sigma}; & 
\nonumber \\
&\pi=-i\,\frac{a\,\sin\theta}{\sqrt{2}\,\Sigma}; \hspace{0.5cm} 
\tau=i\,\frac{a\,\sin\theta}{\sqrt{2}}\,\mu^2; \hspace{0.5cm} 
\alpha=\frac{\cot\theta}{2\,\sqrt{2}}\,\mu^*; \hspace{0.5cm} \beta=-\alpha^* + \tau.& 
\label{coef:spin_esfgen}
\end{eqnarray}

A straightforward substitution of these quantities in the perturbation equation
leads to the following equation for the perturbed scalar of $\Psi_4$:
\begin{equation}
 \frac{1}{2\,\Sigma}\,\left[\square^{\Psi}_{tr} 
+ \square_{\theta\,\varphi} \right]\,{\Psi_4}^{(1)}= 4\,\pi\,T. \label{eq:pertPsi4KS}
\end{equation}
where
\begin{eqnarray}
\square^{\Psi}_{tr}&=&-\left(r^2 + a^2\,\cos^2\theta +2\,M\,r\right)\,\frac{\partial^2}{\partial t^2} +  \Delta\,
\frac{\partial^2}{\partial r^2} + 4\,M\,r\,\frac{\partial^2}{\partial t \partial r}
 + 2\,a\,\frac{\partial^2}{\partial r \partial \varphi} +  \nonumber 
\\ && 2\,\left(2\,r+3\,M + 2\,i\,a\,\cos\theta \right)\,\frac{\partial}{\partial t} 
+ 6\,\left(r - M\right)\,\frac{\partial}{\partial r} + 4 \label{op:Pertrt} \\
\square_{\theta\varphi}&=& \frac{\partial^2}{\partial \theta^2} 
+ \frac1{\sin^2\theta}\,\frac{\partial^2}{\partial \varphi^2} +
\cot\theta\,\frac{\partial}{\partial \theta} -
4\,i\,\frac{\cos\theta}{\sin^2\theta}\,\frac{\partial}{\partial \varphi} -
2\,\frac{1+\cos^2\theta}{\sin^2\theta}. \label{op:Pertthph}
\end{eqnarray}
Notice how remarkable simple turns out to be the operator acting on ${\Psi_4}^{(1)}$,
with an angular operator $\square_{\theta\varphi}$ which is exactly the same as the one
for the Schwarzschild case. In what follows, we will consider only vacuum spacetimes (i.e., $T=0$). 
The explicit form of the operators acting on the source term are
presented in the appendix \ref{appendix1}.

Our operator (\ref{eq:pertPsi4KS}-\ref{op:Pertthph}) has exactly the same form as 
the operator obtained in \cite{Cam01}. However, in their case the operator is acting on the function
$(r-i\,a\,\cos\theta)^4\,{\Psi_4}^{(1)}$ due to a different choice
of the null tetrad. In our case the operator is acting directly on ${\Psi_4}^{(1)}$ since
our tetrad takes the expected form in the spatial asymptotic region.
This implies that the Peeling theorem can be applied, recovering the expected 
decay ${\Psi_4}^{(1)} \sim 1/r$ at far distances. It is
then more convenient to work with the function $\Phi \equiv r\,\,{\Psi_4}^{(1)}$,
which will have a constant behavior in the regions far from the black hole.
The perturbation equation (\ref{eq:pertPsi4KS}) for this quantity takes the form
\begin{equation}
\frac{1}{2\,r\,\Sigma}\,\left[{\square}^{\Phi}_{tr}  + \square_{\theta\,\varphi}
\right]\,\Phi=0. \label{eq:pertPhi1}
\end{equation}
There is only a change in the radial-temporal operator, which now is written as
\begin{eqnarray}
{\square}^{\Phi}_{tr} =&&-\left(r^2 + a^2\,\cos^2\theta +2\,M\,r\right)\,\frac{\partial^2}{\partial t^2} +  \Delta\,
\frac{\partial^2}{\partial r^2} + 4\,M\,r\,\frac{\partial^2}{\partial t \partial r} 
+ 2\,a\,\frac{\partial^2}{\partial r \partial \varphi} +  \nonumber 
\\ && 2\,\left(2\,r+ M + 2\,i\,a\,\cos\theta \right)\,\frac{\partial}{\partial t} 
+ 2\,\left(2\,r - M - \frac{a^2}{r}\right)\,\frac{\partial}{\partial r} -
2\,\frac{a}{r}\,\frac{\partial}{\partial \varphi} +
2\frac{M\,r + a^2}{r^2}
 \label{op:Pert1rt}.
\end{eqnarray}

The main advantage of this perturbation equation with respect to
previous ones found in the literature is that the angular operator (\ref{op:Pertthph}) can be 
expressed in terms of the eth operators \cite{Newm66,Gold67}, just as in the Schwarzschild case
\cite{JC09}:
\begin{equation}
\square_{\theta\varphi}={\bar\eth}_{-1}\,\eth_{-2},
\end{equation}
This implies that the eth operators act initially on a function of spin weight $-2$.
Since the $\eth$ operator raises the spin weight and the ${\bar \eth}$ one lowers it,
their combined action finally give again a function of spin weight
$-2$, so that ${Y_{-2}}^{l,m}$ is an eigenfunction of such operator:
\begin{equation}
\square_{\theta\varphi}\,{Y_{-2}}^{l,m}={\bar\eth}_{-1}\,\eth_{-2}\,{Y_{-2}}^{l,m}
= -\left(l-1\right)\,\left(l+2\right)\,{Y_{-2}}^{l,m}. \label{eq:Ym2}
\end{equation}
In addition, we will take advantage from the fact that 
they are also eigenfunctions of the azimuthal operator, namely
\begin{equation}
\frac{\partial}{\partial
\varphi}\,{Y_{-2}}^{l,m} = i\,m\,{Y_{-2}}^{l,m}. \label{eq:zYm2}
\end{equation}
These results provide us with a natural way to choose the spin-weighted spherical harmonics 
as a basis to generate the space of solutions for the perturbation equation. This approach
is different from the standard one, which started by Teukolsky \cite{Teu73} and have been
used by many others authors. In the standard approach, the spin-weighted spheroidal
harmonic functions are introduced as an eigenfunctions of the
angular operator containing the rotation parameter of the black hole. The spin-weighted
spherical harmonics would correspond to a subset of these functions. The use of this basis allows 
us to know the exact form of the constant that replaces the 
angular part, so that it is no longer an unknown to be determined by the expansions procedures, as
for instance in \cite{Leaver85}.
%
%
%

\section{The one dimensional system of Teukolsky equations}\label{sec:decomposition}

In order to separate the angular dependence we first express the perturbation function $\Phi$ in 
terms of the standard spin-weighted spherical harmonics:
\begin{equation}
\Phi=\sum\limits_{lm}\,R_{l,m}(t,r)\,{Y_{-2}}^{l,m}(\theta, \varphi).
\label{eq:expPhi}
\end{equation}
with $l\geq 2$ because the gravitational waves are quadrupole and higher. By using that
the spherical harmonics are eigenfunction of the angular (\ref{eq:Ym2})
and the azimuthal operator (\ref{eq:zYm2}) we can get almost a
radial-temporal equation, namely
\begin{eqnarray}
&& \sum\limits_{lm}\,{Y_{-2}}^{l,m}\,\left[-\left(r^2 + a^2\,\cos^2\theta
+2\,M\,r\right)\,\frac{\partial^2}{\partial t^2} 
+  \Delta\,
\frac{\partial^2}{\partial r^2} + 4\,M\,r\,\frac{\partial^2}{\partial t \partial r}  
+ 2\,\left(2\,r+ M + 2\,i\,a\,\cos\theta \right)\,\frac{\partial}{\partial t} + \right.
\nonumber 
\\ && \left.  2\,\left(2\,r - M - \frac{a^2}{r} + i\,a\,m\right)\,\frac{\partial}{\partial r} - 2\,i\,m\frac{a}{r} +
2\frac{M\,r + a^2}{r^2} -\left(l-1\right)\,\left(l+2\right) \right]\,R_{l,m}=0.
\end{eqnarray}
There are only left two terms with angular dependence, which are proportional to
$\cos\theta$ and $\cos^2\theta$. Following the usual procedure to obtain an equation
for the coefficients $R_{l,m}$, we multiply the previous equation by the complex
conjugate ${\bar{Y}_{-2}}^{l^\prime,m^\prime}={Y_{2}}^{l^\prime,-m^\prime}$, and integrate
over the angles:
\begin{eqnarray}\label{eqR3}
&& \sum\limits_{lm}\,\oint\,d\Omega\,{Y_{2}}^{l^\prime,-m^\prime}\,{Y_{-2}}^{l,m}\,
\left[-\left(r^2  +2\,M\,r\right)\,\frac{\partial^2}{\partial t^2} +  \Delta\,
\frac{\partial^2}{\partial r^2} + 4\,M\,r\,\frac{\partial^2}{\partial t \partial r}  
+  2\,\left(2\,r+ M  \right)\,\frac{\partial}{\partial t} +  \right.   \nonumber \\
&& \left.  2\,\left(2\,r - M - \frac{a^2}{r} + i\,a\,m \right)\,\frac{\partial}{\partial r} - 2\,i\,m\frac{a}{r} +
2\frac{M\,r + a^2}{r^2} -\left(l-1\right)\,\left(l+2\right)\right]\,R_{l,m}
- 
\nonumber \\
&&a^2\,\oint\,d\Omega\,{Y_{2}}^{l^\prime,-m^\prime}\,{Y_{-2}}^{l,m}\,\cos^2\theta\,\frac{
\partial^2}{\partial t^2}R_{l,m} 
+
4\,i\,a\,\oint\,d\Omega\,{Y_{2}}^{l^\prime,-m^\prime}\,{Y_{-2}}^{l,m}\,\cos\theta\,\frac{
\partial}{\partial t}R_{l,m} =0.
\end{eqnarray}
where we have splitted the sum to isolate the coefficients which include angular
functions. 

By using the orthonormalization relation of the spin-weighted spherical
harmonics, the angular integral of ${Y_{2}}^{l^\prime,-m^\prime}\,{Y_{-2}}^{l,m}$ is reduced 
to $\delta_{l,l^\prime}\,\delta_{m,m^\prime}$. On the other hand, the cosinus appearing on
the last two terms can be written in terms of the spin-weighted spherical harmonics,namely
\begin{equation}
\cos^2\theta=\frac43\,\sqrt{\frac{\pi}5}{Y_{0}}^{2,0} + \frac13; \hspace{1cm} \cos\theta=2\,\sqrt{\frac{\pi}3}{Y_{0}}^{1,0}.
\end{equation}
Substituting these relations in the eq.~(\ref{eqR3}) lead to two angular integrals with
three spin-weighted spherical harmonics. These angular integrals can be solved
by means of the Wigner $3-lm$ symbols (see \cite{Ruiz08} and references therein): 
\begin{eqnarray}
&\oint Y_{2}^{l,-m} \,Y_{-2}^{l^\prime,m^\prime}\,Y_{0}^{1,0}\,
d\Omega=\sqrt{\frac{3}{4\pi}}\,\delta_{m,m^\prime} \left[ B_{l,m}\,\delta_{l,l^\prime-1} +
\frac{2m}{l(l+1)}\,\delta_{l,l^\prime} 
+ B_{l+1,m}\,\delta_{l,l^\prime+1} \right]& \nonumber
\\
&\oint Y_{2}^{l,-m} \,Y_{-2}^{l^\prime,m^\prime}\,Y_{0}^{2,0}\,
d\Omega=\frac34\,\sqrt{\frac{5}{\pi}}\,\delta_{m,m^\prime}\,\left[
A_{l+2,m}\,\delta_{l,l^\prime-2} + 16\,\frac{m}{(l+2)\,l}B_{l+1,m}\,\delta_{l,l^\prime - 1} +
\frac{2\,(l+4)(l-3)(l(l+1)-3m^2)}{3\,(2l+3)(l+1)(l)(2l-1)}\,\delta_{l,l^\prime}
+ \right. & \nonumber \\ & \left. 16\,\frac{m}{(l+1)(l-1)}B_{l,m}\,\delta_{l,l^\prime +1} +
A_{l,m}\,\delta_{l,l^\prime - 2} \right]&
\end{eqnarray}

In this way, it is obtained a 1D coupled system of evolution equations for the gravitational
perturbations in Kerr spacetimes, which can finally be written as
\begin{eqnarray}
&-a^{2}\,A_{l+2,m}\,\partial_{tt}\,R_{l+2,m} - 4\,a\,B_{l+1,m}\,
\left(4\,a\,\frac{m}{\left(l+2\right)\,l}\partial_{tt} -
i\,\partial_{t}\right)\,R_{l+1,m}+\overline{\square}_{l,m}\,R_{l,m} - & \nonumber  \\
&4\,a\,B_{l,m}\,\left(4\,a\,\frac{m}{\left(l+1\right)\,\left(l-1\right)}\partial_{tt} 
- i\,\partial_{t}\right)R_{l-1,m} -
a^{2}\,A_{l,m}\,\partial_{tt}\,R_{l-2,m}=0,&\label{eq0:ev_cou_mod}
\end{eqnarray}
where we have dropped the primes in $l,m$ and we have defined
\begin{eqnarray}
A_{l,m}&=&\frac{1}{l\,(2l-1)\,(l-1)}\sqrt{\frac{(l-2)(l-3)(l+2)(l+1)(l+m)(l+m-1)(l-m)(l-m-1)}{(2l+1)(2l-3)}},\\
B_{l,m}&=&\frac1l\,\sqrt{\frac{(l-2)(l+2)(l+m)(l-m)}{(2l+1)(2l-1)}},\\
\overline{\square}_{l,m}&=&-\left(r^2  +2\,M\,r + \frac{a^2}3\,
\left(1 + 2\,\frac{(l+4)\,(l-3)\,(l\,(l+1)-3m^{2})}{(2l+3)\,(l+1)\,l\,(2l-1)}\right)\right)\,\frac{
\partial^2}{\partial t^2} +  \Delta\,
\frac{\partial^2}{\partial r^2} + 4\,M\,r\,\frac{\partial^2}{\partial t \partial r}  +   \\
&&  2\,\left(2\,r+ M  + 4\,i\,a\,\frac{m}{l(l+1)}\right)\,\frac{\partial}{\partial t} + 
2\,\left(2\,r - M - \frac{a^2}{r} + i\,a\,m \right)\,\frac{\partial}{\partial r} -
2\,i\,m\frac{a}{r} + 2\frac{M\,r + a^2}{r^2} -\left(l-1\right)\,\left(l+2\right).\nonumber
\end{eqnarray}
The angular part dependence has been replaced by a coupling among modes, achieving the goal
of reducing, in the time domain, the perturbation equation in a Kerr background to a
radial-temporal system of (coupled) equations.

The angular part of the gravitational perturbation is described by the spin-weighted
spherical harmonics, while the radial-temporal part is obtained by solving the system of
equations (\ref{eq0:ev_cou_mod}). This system shows the couplings and interdependence of
the modes. The $R_{l,m}$ mode is coupled via the $l$ number with their neighbors,
two above and two below. The modes are not coupled with respect to the
$m-$mode. The coupling coefficients $A_{l,m}$ and $B_{l,m}$ grow with $l$,
indicating an asymmetry on the coupling. They tend to asymptotic finite values which may be 
used to bound the coefficients, $\lim\limits_{l\to \infty}\,A_{l,m}=\frac{1}{4}$ and
$\lim\limits_{l\to \infty}\,B_{l,m}=\frac12$. They depend symmetrically on $m$ and grow with its
modulus. Finally, we see that these couplings are determined by the rotational parameter $a$ of the
black hole, and vanish in the Schwarzschild case.

Another interesting property of this system is that they are coupled through the
temporal derivatives of the modes, while the radial dependence for a given $(l,m)$ mode 
is determined by the box operator $\overline{\square}_{l,m}$ acting on it. This means that
the asymptotic behavior of the $R_{l,m}$ mode is determined by the box operator,
since the couplings are lost in that limit. Indeed, we find that solving the equation
\begin{equation}
\overline{\square}_{l,m} \,R_{l,m}=0,
\label{eq:boxzero}
\end{equation}
is enough to identify the main properties of the $R_{l,m}$ mode in regions far from the black
hole. The numerical evolutions described in section \ref{sec:evolution} will show that solving 
this equation already gives a good description of the solution.This will be relevant in the study of
the properties of late time behavior of the gravitational perturbation as tails (see \cite{Zeng08,
Zeng08a} for a review on the subject), or in the fast and efficient generation of waveforms. Notice
that in the Schwarzschild case (i.e., $a=0$)
the modes decouple and we are left with only the equation (\ref{eq:boxzero}) for each mode.

The system (\ref{eq0:ev_cou_mod}) has another peculiarities. On one hand, since the gravitational
waves only have quadrupole and higher contributions, the radial-temporal modes $\{R_{1,m},
R_{0,m}\}$ are zero by construction and the first non-trivial equation would
correspond to $l=2$, namely
\begin{eqnarray}
-a^{2}\,A_{4,m}\,\partial_{tt}\,R_{4,m} - 4\,a\,B_{3,m}\,
\left(\frac{a\,m}{2}\partial_{tt} -
i\,\partial_{t}\right)\,R_{3,m}+\overline{\square}_{2,m}\,R_{2,m} =0 ~~.\label{eq:l2}
\end{eqnarray}
On the other hand, the coupled system contains an infinity set of equations involving
all possible values of $l\geq 2$. In order to solve it, the expansion (\ref{eq:expPhi}) must be 
truncated at some order $l_{max}$. From there on it is assumed that $R_{l>l_{max},m}=0$, which
prevents a overdeterminated system. The last equation of the series would look like:
\begin{eqnarray}\label{eq:lmax}
 \overline{\square}_{l_{max},m}\,R_{l_{max},m} -
4\,a\,B_{l_{max},m}\,\left(4\,a\,\frac{m}{\left(l_{max}+1\right)\,\left(l_{max}-1\right)}\partial_{tt} 
- i\,\partial_{t}\right)R_{l_{max}-1,m} -
a^{2}\,A_{l_{max},m}\,\partial_{tt}\,R_{l_{max}-2,m}=0~~.
\end{eqnarray}
The modes $l=3$ and $l=l_{max-1}$ are also specials, since there still appear zero contributions.
The 
rest of the modes in the system of evolution equations (\ref{eq0:ev_cou_mod}) are coupled by the
next two values of the parameter $l$, above and below a given one.

%
%

\section{The first order evolution equations}\label{sec:firstorder}

In this section we will present two different formulations of the Teukolsky equations
suitable for numerical evolutions. The first one is the first order reduction of the
$1+1$ coupled system (\ref{eq0:ev_cou_mod}) described in the previous section, where
several $l-$modes will be taken into account. In order to check the results, we will
compare with a standard $2+1$ formulation to solve the gravitational perturbation equation. This
second approach uses a different base of function to describe the angular dependence,
leading to a two dimensional uncoupled system in the coordinates $(r,\theta)$.

\subsection{The coupled one-dimensional system}

Since the Weyl scalar is complex, we first split the radial function into a real and imaginary part, that is,
\begin{equation}
R_{l,m}={}^+R_{l,m} + i\,{}^-R_{l,m} ~~.
\end{equation}
so that the eq. (\ref{eq0:ev_cou_mod}) decouples into the following second order system of equations:
\begin{eqnarray}
&-a^{2}\,A_{l+2,m}\,\partial_{tt}\,{}^{\pm}R_{l+2,m} - 4\,a\,B_{l+1,m}\,
\left(4\,a\,\frac{m}{\left(l+2\right)\,l}\partial_{tt}\,{}^{\pm}R_{l+1,m} \pm
\partial_{t}\,{}^{\mp}R_{l+1,m}\right)+{}_r\overline{\square}_{l,m}\,{}^{\pm}R_{l,m} -&  
\label{eq1:ev_cou_mod} \\
&\mp 2\,a\,m\,\left(\frac4{\left(l+1\right)\,l}\partial_t + \partial_r - 
\frac1r \right){}^{\mp}R_{l,m} -
4\,a\,B_{l,m}\,\left(4\,a\,\frac{m}{\left(l+1\right)\,\left(l-1\right)}\partial_{tt}\,{}^{\pm}
R_{l-1,m} \pm  \partial_{t}\,{}^{\mp}R_{l-1,m}\right) -
a^{2}\,A_{l,m}\,\partial_{tt}\,{}^{\pm}R_{l-2,m}=0,&\nonumber
\end{eqnarray}
where we have defined the operator ${}_r\overline{\square}_{l,m}$ as
\begin{eqnarray}
{}_r\overline{\square}_{l,m}&=&-\left(r^2  +2\,M\,r + \frac{a^2}3\,
\left(1 +
2\,\frac{(l+4)\,(l-3)\,(l\,(l+1)-3m^{2})}{(2l+3)\,(l+1)\,l\,(2l-1)}\right)\right)\,\frac{
\partial^2}{\partial t^2} +  \Delta\,
\frac{\partial^2}{\partial r^2} + 4\,M\,r\,\frac{\partial^2}{\partial t \partial r}  +   \\
&&  2\,\left(2\,r+ M\right)\,\frac{\partial}{\partial t} +  2\,
\left(2\,r - M - \frac{a^2}{r} \right)\,\frac{\partial}{\partial r} +
2\frac{M\,r + a^2}{r^2} -\left(l-1\right)\,\left(l+2\right).\nonumber
\end{eqnarray}

The reduction from second order in space to first order can be achieved by defining the functions
\begin{eqnarray}
{}^{\pm}\Psi_{l,m} &=& \partial_r\,{}^{\pm}R_{l,m},  \\
{}^{\pm}\Pi_{l,m} &=& \partial_t\,{}^{\pm}R_{l,m} + \beta\,{}^{\pm}\Psi_{l,m}, 
\end{eqnarray}
%

For the Schwarzschild case $\beta$ is directly the shift vector and the above definition
is useful to align the wave speed to the light cones. For the Kerr solution this is not
the case but we explore the effect of varying it, the resultant wave
form is insensitive to our choice and we set the easiest value of $\beta=1$, but we leave
it as a free parameter in the equations.

From these
definitions we then obtain
\begin{eqnarray}
\partial_t\,{}^{\pm}R_{l,m} &=&{}^{\pm}\Pi_{l,m} - \beta\,{}^{\pm}\Psi_{l,m}, \\
\partial_t\,{}^{\pm}\Psi_{l,m} &=&\partial_r\,\left({}^{\pm}\Pi_{l,m} -
\beta\,{}^{\pm}\Psi_{l,m}\right), \\
\partial_{tt}\,{}^{\pm}R_{l,m} &=&\partial_t\,{}^{\pm}\Pi_{l,m} -
\beta\,\partial_r\,\left({}^{\pm}\Pi_{l,m} - \beta\,{}^{\pm}\Psi_{l,m}\right).
\end{eqnarray}

Using this last equation we obtain the following first order equations:
\begin{eqnarray}
&\left[(T_1,T_2,T_3,T_4,T_5)\partial_{t} + (c_1,c_3,c_5,c_7,c_9,)\partial_{r}\right]\,
\left(\begin{array}{l l l l l}
{}^{\pm}\Pi_{l+2,m}\\ {}^{\pm}\Pi_{l+1,m} \\ {}^{\pm}\Pi_{l,m} \\ {}^{\pm}\Pi_{l-1,m} \\
{}^{\pm}\Pi_{l-2,m}
\end{array}\right) + \big[(c_2,c_4,c_6,c_8,c_{10})\partial_{r} +
(S_1,S_3,S_7,S_{12},S_{14}) \big]\,\left(\begin{array}{l l l l l}
{}^{\pm}\Psi_{l+2,m}\\ {}^{\pm}\Psi_{l+1,m} \\ {}^{\pm}\Psi_{l,m} \\ {}^{\pm}\Psi_{l-1,m} \\
{}^{\pm}\Psi_{l-2,m}
\end{array}\right)& \nonumber \\
& + (0,S_2,S_6,S_{11},0)\, \left(\begin{array}{l l l l l}
{}^{\mp}\Pi_{l+2,m}\\ {}^{\mp}\Pi_{l+1,m} \\ {}^{\mp}\Pi_{l,m} \\ {}^{\mp}\Pi_{l-1,m} \\
{}^{\mp}\Pi_{l-2,m}
\end{array}\right) +
(0,S_4,S_8,S_{13},0)\, \left(\begin{array}{l l l l l}
{}^{\mp}\Psi_{l+2,m}\\ {}^{\mp}\Psi_{l+1,m} \\ {}^{\mp}\Psi_{l,m} \\ {}^{\mp}\Psi_{l-1,m} \\
{}^{\mp}\Psi_{l-2,m}
\end{array}\right) + S_5\,{}^{\pm}\Pi_{l,m}+S_9\,{}^{\pm}R_{l,m}+S_{10}\,{}^{\mp}R_{l,m}=0,&
\label{eq:pert1d}
\end{eqnarray}
which are the final first order equations. The coefficients are defined as
\begin{eqnarray}
&T_1=-a^{2}\,A_{l+2,m}, \,\,\, T_2=-16\,a^2\,B_{l+1,m}\,\frac{m}{\left(l+2\right)\,l}, \,\,\, T_3
=-\left(r^2  +2\,M\,r + \frac{a^2}3\,\left(1 + 2\,\frac{(l+4)\,(l-3)\,(l\,(l+1)-3m^{2})}{(2l+3)\,(l+1)\,l\,(2l-1)}\right)\right), 
& \nonumber \\
&T_4=-16\,a^2\,B_{l,m}\,\frac{m}{\left(l+1\right)\,\left(l-1\right)}, \,\,\, T_5=-a^{2}\,A_{l,m}& \nonumber \\
&c_1=-T_1\,\beta, \,\,\, c_2=T_1\,\beta^2, \,\,\, c_3=-T_2\,\beta, \,\,\,
c_4=T_2\,\beta^2, \,\,\,
 c_5=T_3\,\left(4\,M\,r - \beta\right), \,\,\, c_6=T_3\,\left(\beta^2 - 4\,M\,r\,\beta
+\Delta\right)&,  \nonumber \\
&c_7=-T_4\,\beta, \,\,\, c_8=T_4\,\beta^2, \,\,\, c_9=-T_5\,\beta, \,\,\,
c_{10}=T_5\,\beta^2,  \nonumber \\ &S_1=T_1\,\beta\,\partial_r\beta, \,\,\, S_2=\mp
4\,a\,B_{l+1,m}, \,\,\, S_3=T_2\,\beta\,\partial_r\beta, \,\,\, S_4=\pm
4\,a\,B_{l+1,m}\,\beta,  \,\,\, S_5=2\,\left(2\,r + M\right)\,T_3,& \nonumber \\
&S_6=\mp 8\,a\,\frac{m}{\left(l+1\right)\,l}\,T_3, \,\,\, 
S_7=\left(\left(\beta- 4\,M\,r\right)\,\partial_r\beta - 2\,\left(2\,r + M\right)\,\beta + 
2\,\left(2\,r - M - \frac{a^2}r\right)\right)\,T_3,& \nonumber \\
&S_8=\mp 2\,a\,m\,\left(1 - \beta\,\frac4{\left(l+1\right)\,l}\right)\,T_3, \,\,\, 
S_9=\left(2\frac{M\,r + a^2}{r^2} -\left(l-1\right)\,\left(l+2\right)\right)\,T_3, \,\,\, 
S_{10}=\pm 2\,m\frac{a}{r}\,T_3, & \nonumber \\
&S_{11}=\mp 4\,a\,B_{l,m}, \,\,\, S_{12}=T_4\,\beta\,\partial_r\beta, \,\,\, 
S_{13}=\pm 4\,a\,B_{l,m}\,\beta, \,\,\, S_{14}=T_5\,\beta\,\partial_r\beta. & \nonumber
\end{eqnarray}
By reducing the system to first order, we have enlarged our set of variables for each mode
$(l,m)$ from 2 (real and imaginary parts ${}^{\pm}R_{l,m}$) to 6 ($\{{}^{\pm}R_{l,m},{}^{\pm}\Pi_{lm}$
${}^{\pm}\Psi_{lm} \}$). In order to solve this system of coupled equations
we write it as a matrix equation, $M\partial_t\,\vec{u} + A\,\partial_r\,\vec{u} +B\,\vec{u} =0$.
The 
matrix $M$ has to be inverted in order to obtain the evolution equation for each variable.

\subsection{The two-dimensional system}

We will follow here the approach given by Krivan et al.\cite{Kri97} but for the 
Teukolsky function $\Phi=r\,{\Psi_4}^{(1)}$ instead of $(r-ia\cos\theta)^4\,{\Psi_4}^{(1)}$, which
will introduce some differences on the final equations. Assuming that the solution 
to eq. (\ref{eq:pertPhi1}) can be expressed in terms of the azimuthal modes, that is,
\begin{equation}
\Phi =\sum\limits_{m}\,e^{i\,m\,\varphi}\,S_m(r,\theta,t), 
\end{equation}
it is obtained, for each azimuthal mode $m$, the following perturbation equation: 
\begin{eqnarray}
\label{eq:Sm}
&&\left(-\left(r^2 + a^2\,\cos^2\theta +2\,M\,r\right)\,\frac{\partial^2}{\partial t^2} +
 \Delta\,
\frac{\partial^2}{\partial r^2} + 4\,M\,r\,\frac{\partial^2}{\partial t \partial r} + 
\frac{\partial^2}{\partial \theta^2}  + 2\,\left(2\,r+ M + i\,a\,\cos\theta \right)\,\frac{\partial}{\partial t} 
+  \right.  
\\ && \left. 2\,\left(2\,r - M - \frac{a^2}{r} + i\,a\,m\right)\,\frac{\partial}{\partial r} +  
\cot\theta\,\frac{\partial}{\partial \theta} +
2\left(\frac{M\,r + a^2}{r^2} -i\,\frac{a\,m}{r}\right)  -\frac{m^2 - 4\,m\,\cos\theta 
+ 2\,\left(1+\cos^2\theta\right)}{\sin^2\theta} \right)\,S_m=0.\nonumber
\end{eqnarray}
The function $S_m$ is complex, so we separate it in its real and imaginary parts
\begin{equation}
 S_{m}=S_{+}+iS_{-},
\end{equation}
and obtain two sets of equations
\begin{eqnarray*}
 P_{+}S_{+}-P_{-}S_{-}=0,\\
 P_{+}S_{-}+P_{-}S_{+}=0.
\end{eqnarray*}
where $P_{+}$ and $P_{-}$ refers to the real and imaginary parts of the operator in
equation (\ref{eq:Sm}).As we did before, a first order reduction can be obtained by
defining the variables $\Pi_\pm$ as 
\begin{equation}
\Pi_\pm = \partial_{t}S_\pm+b\,\partial_{r}S_\pm,
\end{equation}
where 
\begin{equation}
b=\frac{1}{\Sigma+2Mr}\left(-2Mr+ \sqrt{ (2Mr)^{2}+\Delta(\Sigma+2Mr) } \right) .    
\end{equation}
In terms of the new variables $\Pi_{+}, \Pi_{-}$ the perturbation equation can be
rewritten as:
\begin{eqnarray}
 \partial_{t}\Pi_{\pm}&=&c_{1}\partial_{r}\Pi_{\pm}+c_{2}\partial_{r}S_{+}\pm
c_{3}\partial_{r}S_{\mp}
- c_{4}\Pi_{\pm}\pm c_{5}\Pi_{\mp}-c_{6}S_{\pm}\pm c_{7}S_{\mp}+{\cal O}_{\theta,\phi}S_{\pm},\\
 \partial_{t}S_{\pm}&=&\Pi_{\pm}-b\,\partial_{r}S_{\pm},
 \label{eq:2dmatriz}
\end{eqnarray}
where now we have defined the following functions
\begin{eqnarray*}
c_{1}&=& -\frac{1}{\Sigma+2Mr}\left(2Mr+\sqrt( (2Mr)^{2}+\Delta(\Sigma+2Mr) ) \right) \\
c_{2}&=& bc_{4}+\frac{2(2r^2-Mr-a^2)}{(\Sigma+2Mr)r}+c_{1} \frac{d}{d\,r}b\\
c_{3}&=&-\frac{2am}{\Sigma+2Mr}-c_{5}b\\
c_{4}&=&-\frac{2(2r+M)}{\Sigma+2Mr}\\
c_{5}&=& -\frac{4a\cos\theta}{\Sigma+2Mr} \\
c_{6}&=&\frac{r^2m^2\csc^2\theta-2(Mr-a^2)}{r^2(\Sigma+2Mr)}+\frac{2(1+\cos^2\theta)-4m\cos\theta}{\sin^2\theta(\Sigma+2Mr)} \\
{\cal O}_{\theta \,\phi} &=&
\frac{1}{\Sigma+2Mr}\partial_{\theta\theta}+\frac{\cot\theta}{\Sigma+2Mr}\partial_{\theta
}
\end{eqnarray*}
Since we are using the Kerr-Schild coordinates, there is no need to introduce 
the tortoise coordinate to avoid singularities, as our chart is regular at the apparent horizon.
Our equation (\ref{eq:Sm}) is similar in form at the one given by Campanelli et al. \cite{Cam01}.

%
%

\section{Numerical simulations}\label{sec:evolution}

In this section we will evolve and analyze the gravitational
response of the background spacetime for different initial perturbations. First,
we will compute the truncated solution of the one-dimensional coupled system
(\ref{eq:pert1d}) for different $l_{max}=2,3$ and $4$, leading to a system with
$6,12$ and $18$ unknowns respectively. These solutions will be compared with the solution
from the usual two-dimensional perturbation equation (\ref{eq:2dmatriz}).
This procedure will not only prove the validity of the 1d description, 
but also will allow us to better understand the role 
of each mode as a function of the mode of the initial perturbation.
Next, we will solve only for $l_{max}=2$ (i.e., the first order reduction of the harmonic 
equation (\ref{eq:boxzero})), comparing again with the solution from the two-dimensional system. As
we will show, solving this single decoupled equation for each mode is an efficient
way to describe quite accurately the gravitational perturbations.

The numerical simulations have been performed with an extension of the code used in \cite{JC09} for 
solving the perturbation equation in a Schwarzschild background. The equations are evolved by means
of the Method Of Lines, with a third order Runge-Kutta for the time evolution and a fourth order
differences to approximate the radial derivatives. A small amount of Kreiss-Oliger dissipation is
added for numerical stability reasons.

The reliability of the numerical implementation and the accuracy of the solutions was
checked by performing several evolutions with different resolutions in order to 
determine the convergence of our results. We conclude that the solution shows a
third order convergence, although higher resolutions are needed (i.e.,$\Delta r \sim 0.04$
for the finest grid) to keep this convergence order in the cases with high spins of the
black hole (i.e., $a=0.999$).

\subsection{Initial data with a l=2 mode}

We study the response of a black hole with mass $M=1$ to some initial infalling gravitational
perturbation, for two different values of the rotational parameter $a=0.2$ and $a=0.9$.
The angular dependence of the initial data is described by a single spin-weighted
spherical harmonic mode, either with $(l=2,m=0)$ or $(l=2,m=2)$. The radial part has a Gaussian 
profile of the form $g(r)=A\exp(-(r-r_0)/\sigma^{2})$, with an amplitude, center and width
respectively of $A = 5\times 10^{-3}$, $r_0 = 20M$, $\sigma = 0.5$. Its time derivative is assumed
to be zero initially. This initial perturbation is evolved with the 1d harmonic decomposition,
solving the equation (\ref{eq:l2}) with $R_{2,m}(r,t=0)=g(r)$.
The one dimensional coupled system is evolved with $l_{max}=2,3$ and $4$ in order
to calculate how fast the truncated solution converges to the complete one.

In order to compare with the $2d$ approach, we have to write the corresponding initial data
on this basis. For instance, in the case of the mode $(l=2,m=0)$, that is
\begin{equation}
 S_{+}(r,\theta,t=0)=g(r)\,{Y_{-2}}^{2 0}(\theta),
\label{eq:Sminit}
\end{equation}
with
\begin{equation}
 {Y_{-2}}^{2 0}(\theta)= \frac{1}{8}\sqrt{\frac{30}{\pi}}\sin^{2}(\theta)~.
\end{equation}
We solve in a $2-$dimensional grid the equation (\ref{eq:2dmatriz}), with the
initial data given above (\ref{eq:Sminit}). In order to compare the solutions,
we have to project the solution from the 2d system onto a spin weighted
spherical harmonic basis, as long as we will compare the radial 
part of the wave form. In this way, we use the projection:
\begin{equation}
 {}^{2d}R_{2,0}=\oint S_{+}(r, \theta){\bar{Y}_{-2}}^{2 0}(\theta)\,d\Omega,
\end{equation}
and then compare both signals, namely, the one produced by the 1d harmonic decomposition and 
the one obtained by means of the standard 2d evolution. 

\begin{figure}
\centering
\includegraphics[scale=0.4,angle=0]{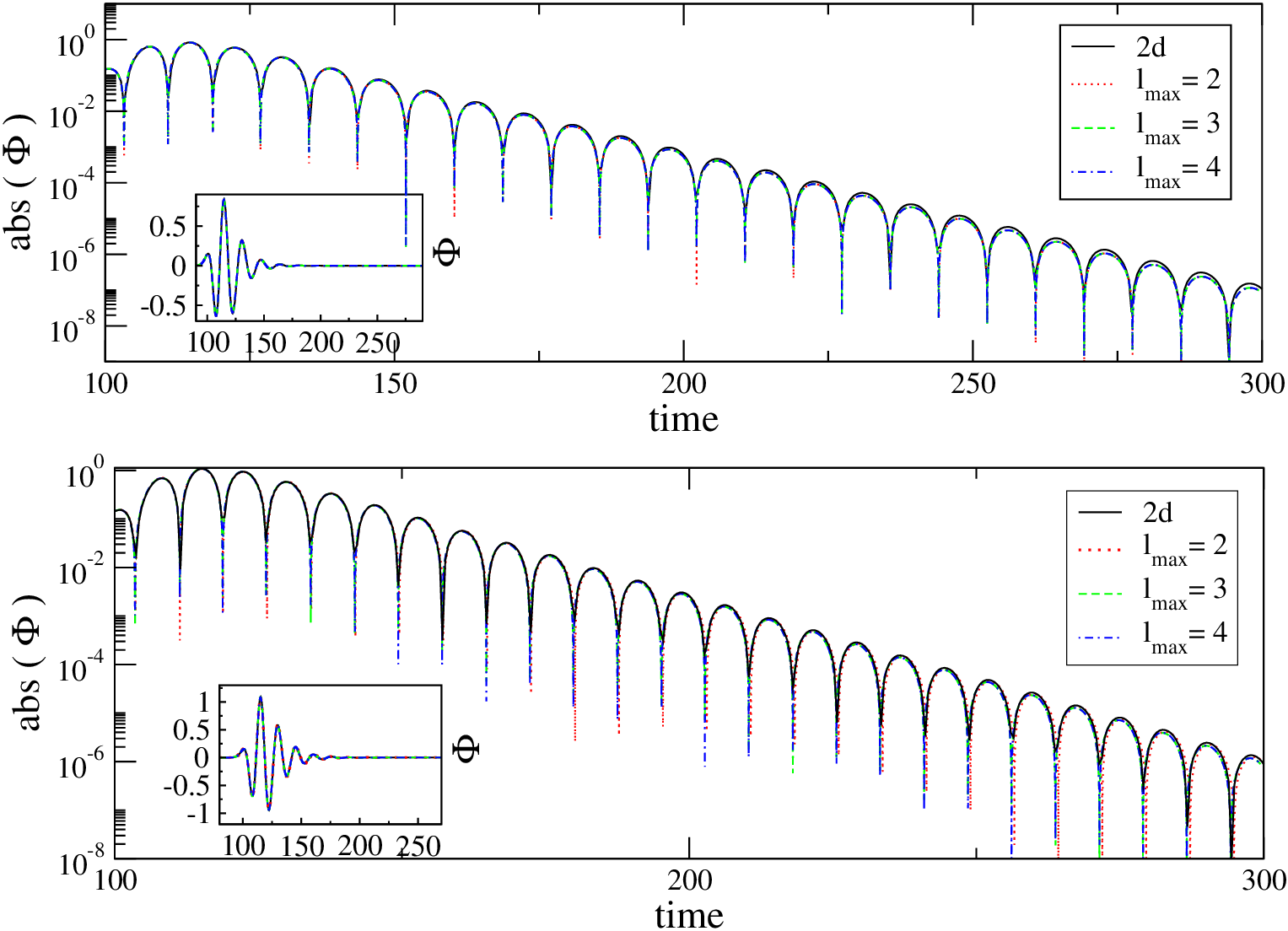}
\caption{We present two values for the rotational parameter of the black hole $a=0.2$ and
$a=0.9$ for an initial perturbation described by the $l=2,m=0$ mode. In the first panel we show the
waveform measured by one observer located at $r=70M$, in
the second the same signal but in logarithmic
scale for $a=0.2$. In the second row the waveform obtained for $a=0.9$ }
\label{fig:R20}
\end{figure}

Our results for the $(l=2,m=0)$ mode are shown in Fig.~\ref{fig:R20},
where it is displayed the signal measured by an observer located at $r=70M$ for 
different $l_{max}$ and the corresponding projection of the $2d$ approach. There is a
remarkable agreement in all the waveforms, showing that the results
with $l_{max}=2$ are very similar to the ones with $l_{max}=3$ and $l_{max}=4$ and to the
$2d$ system. This means that the truncated series converge very fast and that the influence 
of the coupled neighbor modes is pretty small. Since solving equation (\ref{eq:l2}) is much less
expensive than solving the 2d code, this may allow us to use a finer grid with more resolution to
describe the waveform.

\begin{figure}
\centering
\includegraphics[scale=0.4,angle=0]{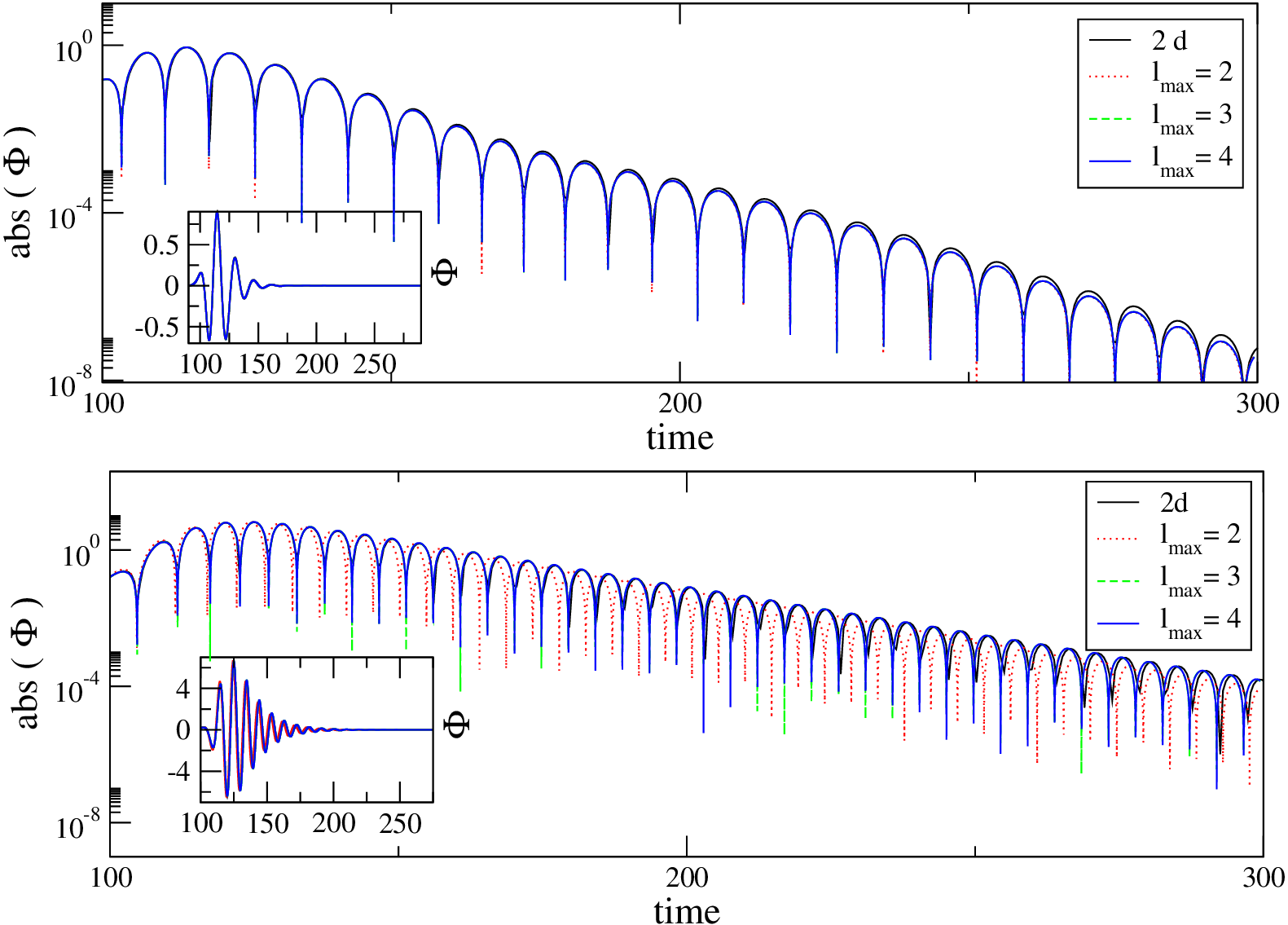}
\caption{As in the previous figure, but for an initial perturbation described by the
$l=2,m=2$ mode}
\label{fig:R22a9}
\end{figure}

Our next example is to consider a initial perturbation with the same radial profile but with
an angular dependence given by the $(l=2,m=2)$ spin-weighted spherical harmonic mode. Notice
that this is an extreme case since the coupling of modes depends strongly 
on $m$. The results are displayed in figure~(\ref{fig:R22a9}). In this case, there is a
noticeable
difference among the results obtained in the 1d harmonic decomposition, depending on the
number of variables we evolve, that is, on the modes that we let to be awaken. 
The contribution of the higher modes is evident and, as we increase number of
modes that we let to be awaken, we get closer to the signal of the $2d$ approach.

The results presented in this subsection not only indicate that the truncated
solution of the 1d coupled system describes accurately the waveforms of a gravitational 
perturbation, but also allow us to conjecture that the coupling among the modes is mainly determined
not so much by the value of the rotation parameter of the black hole, but by the value of the
azimuthal mode number $m$.
 
\subsection{Initial data with two modes with different l}
Next we will consider an initial perturbation whose angular dependence is
described by a combination of the two modes $l=2$ and $l=3$. The problem
is made more challenging  by increasing the black hole rotational parameter to $a=0.999$.
In this case we evolve only within the 1d harmonic decomposition (\ref{eq:boxzero})
for each mode separately. Since any
perturbation can be
decomposed in spin weighted spherical harmonics, this example gives us some
insight about the general behavior on the interaction of modes initially awaken.
The initial data is given by the two modes describe by the same radial function
$R_{2,0} = R_{3,0} = A\exp(-(r-r_0)/\sigma^{2})$, with
the time derivative of both modes set to zero initially.

\begin{figure}
\centering
\includegraphics[scale=0.4,angle=0]{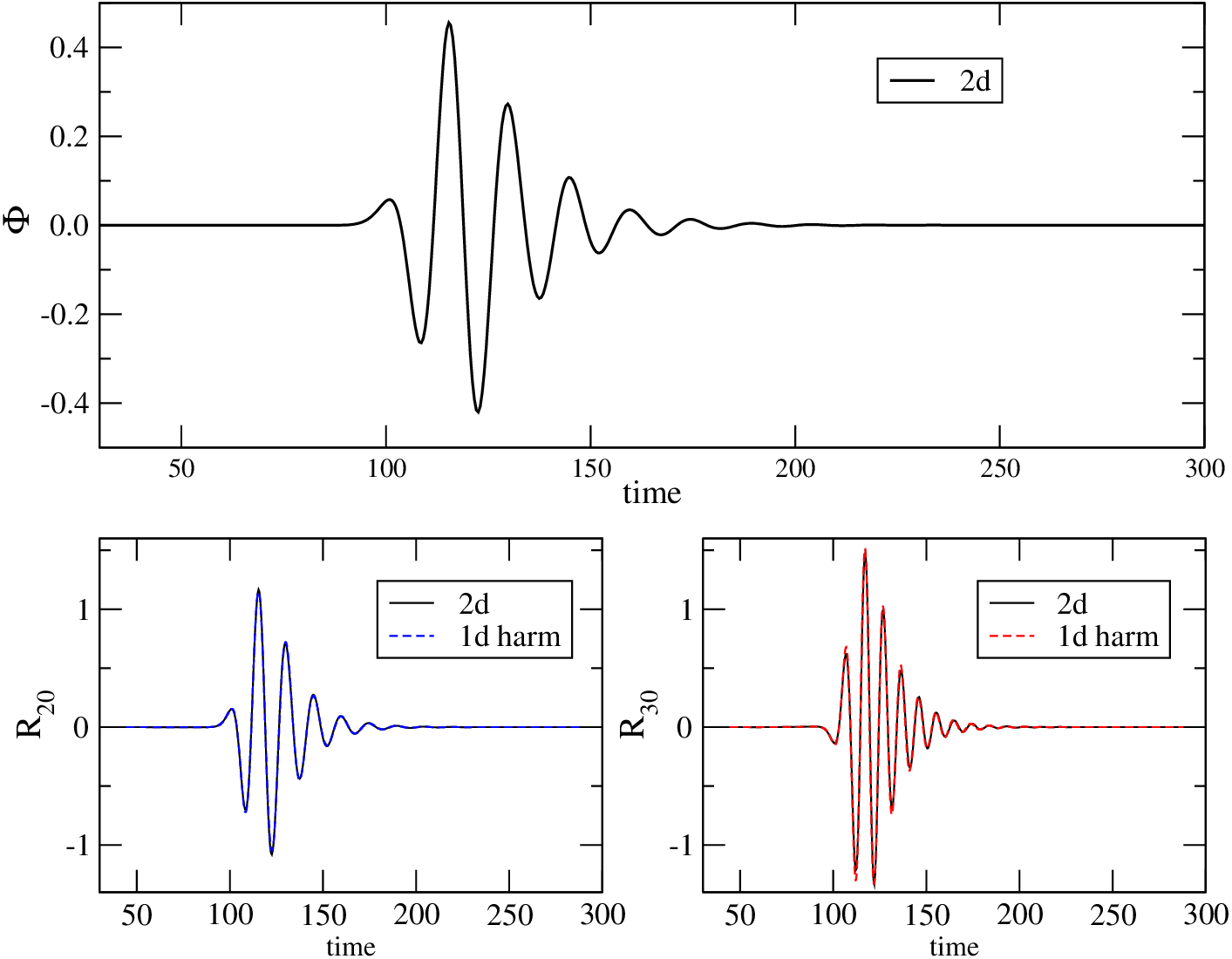}
\caption{It is shown in first panel the signal produced by the axial code,
measured by an observer located at $r=80M$ in the equatorial plane $\theta=\pi/2$,
and the rotational parameter of the black hole is $a=0.999$. In the panel below it is shown a
superposition of the projection of that signal in the $Y_{20}$ and $Y_{30}$ mode 
and the ones obtained by the 1d harmonic decomposition.}
\label{fig:2modes}
\end{figure}

For the 2d approach we have the linear superposition of those modes 
\begin{equation}
S_{+}(r,\theta,t=0)=g(r)\left({{Y}_{-2}}^{2 0}(\theta)+ {{Y}_{-2}}^{3 0}(\theta)
\right).
\end{equation}
As before, we make the projection of the final signal in order to compare both
approaches. The results are plotted in Fig.~(\ref{fig:2modes}).
Notice that, despite the fact that two modes are initially present, we still are
able to describe correctly their evolution by means of the 1d harmonic decomposition, showing 
that it could be used instead of the usual procedure of solving the two dimensional
description for $S_{m}$, even for very large value of the rotational parameter of the black hole.
%

%
%

\section{Conclusions}\label{sec:conclusion}

We have described the perturbation equation for a Kerr black hole in Kerr-Schild
coordinates. By finding a suitable tetrad, we have been able to obtain a perturbation equation
the Weyl scalar $\Psi_4$, which is amiable to be described by the spin weighted spherical
harmonics instead of the spheroidal ones used in other previous works. This fact allowed
us to use the well known properties of the spin weighted spherical 
harmonics for separating the angular dependence from the temporal-radial one. Finally,
we obtained a system of equations on the coupled coefficients of the 
perturbation equation. This coupling is given by the closest modes, two above and two
below (in terms of $l$), and there is not an explicit mixture with respect to the mode number $m$.
Achieved in this way, a separable description for the gravitational perturbation in the Kerr
background, in the time domain.

Once we obtained this system of equations, we proceed to test its actual behavior by making several
numerical codes. We defined new variables that allowed us to
re-write the coupled evolution equation as a first order system. We  solve it numerically by
splitting the real and imaginary parts. We compare the results obtained by this method with the
standard one, writing the perturbation equation by a decomposition in term of the axial modes only.
When we compare our results with the coupled modes with the axial one, we found that the coupling
manifest itself in an stronger way by the mode number $m$, rather than by the values of the angular
parameter of the black hole. Indeed, in the examples that we considered, we saw that the wave
form obtained by solving simply the evolution equation for the corresponding mode without
considering the influences of the neighborhood modes coincides with great accuracy with the wave
form obtained with the axial evolution.
 We also showed that the 
QNM frequencies of the ring down phase of the wave form, agree very well with the known results.

In this way we have found a new, suitable approach that allows to generate accurate wave
forms in a Kerr background in efficient way.
\section*{Acknowledgments}
DN acknowledges DAAD and DGAPA-UNAM grants for
partial support, as well to Lidia Rojas for her support during the elaboration of the
present work. JCD acknowledges CONACyT support. 

\appendix

\section{Source terms}\label{appendix1}

The explicit form of the operators acting on the perturbed stress energy tensor. 

\begin{equation}
T ={\cal{T}}^{k\,k}\,{T^{(1)}}_{k\,k} + {\cal{T}}^{k\,m^*}\,{T^{(1)}}_{k\,m^*} +
{\cal{T}}^{m^*\,m^*}\,{T^{(1)}}_{m^*\,m^*}, \label{eq:T4g}
\end{equation}
with
\begin{eqnarray}
{\cal{T}}^{k\,k}&=&-\left({\bf \delta^*} + \eta_{12}\, \left(3 \alpha + \beta^* + 4\,\pi
- \tau^* \right)\right)\,\left({\bf \delta^*} +\eta_{12}\,\left(2\alpha+2\beta^* -\tau^*
\right)\right), \nonumber \\
{\cal{T}}^{k\,m^*}&=&\left({\bf \Delta} + \eta_{12}\, \left(4\,\mu + \mu^* + 3 \gamma
-\gamma^* \right)\right)\, \left({\bf \delta^*}+2\,\eta_{12}\,\left(\alpha -\tau^*
\right)\right) + \left({\bf \delta^*} + \eta_{12}\, \left(3 \alpha + \beta^* + 4\,\pi -
\tau^* \right)\right)\,\left({\bf \Delta} +  2\eta_{12}\,\left(\mu^* + \gamma
\right)\right), \nonumber \\
{\cal{T}}^{m^*\,m^*}&=&-\left({\bf \Delta} + \eta_{12}\, \left(4\,\mu + \mu^* + 3 \gamma
-\gamma^* \right)\right)\,\left({\bf \Delta}
+\eta_{12}\,\left(\mu^*+2\gamma-2\gamma^*\right)\right). \label{ops:mat_gen}
\end{eqnarray}
We recall that the ${T^{(1)}}_{ab}$ are the projections of the perturbed source term,
${T^{(1)}}_{\mu\,\nu}$ on the null tetrad.

In the Kerr background, with our choice of coordinates and null tetrad, the operators of
$T$, defined by Eqs.~(\ref{eq:T4g},\ref{ops:mat_gen}), have the following explicit
form:
\begin{eqnarray}
{\cal{T}}^{k\,k}&=&-\frac{{\mu^*}^2}{2}\,\left[{\bar\eth}_{-1}\,{\bar\eth}_0
-a^2\,\sin^2\theta\,
\left(\frac{\partial^2}{\partial t^2} - 2\,\left(2\,\mu +
\mu^*\right)\,\left(\frac{\partial}{\partial t} - \mu^*\,\right)\right) -
2\,i\,a\,\sin\theta\,\left(\frac{\partial}{\partial t} -
2\,\mu - \mu^*\right)\,{\bar \eth}_0 \right], \nonumber\\
{\cal{T}}^{k\,m^*}&=&-\sqrt{2}\,\mu^*\,\left[ \left(\frac{\partial}{\partial
t}-\frac{\partial}{\partial r} - 2\,\mu -
\mu^*\right){\bar\eth}_{-1} + i\,a\,\sin\theta\,\left(\frac{\partial^2}{
\partial t^2} - \frac{\partial^2}{\partial t \partial r} -
4\,\mu\,\frac{\partial}{\partial t}
+ 2\,\left(2\,\mu-\mu^*\right)\,\frac{\partial}{\partial r} + 2\,{\mu^*}^2\right) 
\right], \nonumber\\
{\cal{T}}^{m^*\,m^*}&=&-\left[\frac{\partial^2}{\partial
t^2} - 2\,\frac{\partial^2}{\partial t \partial r} + \frac{\partial^2}{\partial r^2} -
2\,\left(2\,\mu+\mu^*\right)\,\left(\frac{\partial}{\partial t} -
\frac{\partial}{\partial r}\right) + 4\,\mu\,\mu^* \right], \label{ops_mat}
\end{eqnarray}
which trivially reduce to the corresponding expressions in the Schwarzschild case,
\cite{JC09}.

\section{Scalar field}

As mentioned in the text, a similar procedure can be use to deal with the motion equation for a
massless scalar field, which satisfies the corresponding Klein-Gordon equation,
$\nabla^\alpha\,\nabla_\alpha\,\Phi=4\,\pi\,T$. Indeed, in the Kerr background, given by the line
element Eq.~(\ref{eq:lel_kerr}), the Klein-Gordon equation takes the explicit form:
\begin{equation}
-\frac{1}{r\,\Sigma}\,\left[\square 
+ \square_{\theta\,\varphi} \right]\,\phi=4\,\pi\,T. \label{eq:phiKS}
\end{equation}
where we have defined $\phi=r\,\Phi$, and the operators:
\begin{eqnarray}
\square&=&-\left(r^2 + a^2\,\cos^2\theta +2\,M\,r\right)\,\frac{\partial^2}{\partial t^2} + 
\Delta\,
\frac{\partial^2}{\partial r^2} + 4\,M\,r\,\frac{\partial^2}{\partial t \partial r}
 + 2\,a\,\frac{\partial^2}{\partial r \partial \varphi}   \nonumber 
\\ && -2\,M\,\frac{\partial}{\partial t} 
+ 2\,\left(M-\frac{a^2}{r}\right)\,\frac{\partial}{\partial r} + 2\,\frac{M\,r - a^2}{r^2}
\label{op:Pertrt_esc} \\
\square_{\theta\varphi}&=& \frac{\partial^2}{\partial \theta^2} 
+ \frac1{\sin^2\theta}\,\frac{\partial^2}{\partial \varphi^2} +
\cot\theta\,\frac{\partial}{\partial \theta} -
2\,\frac{a}{r}\,\frac{\partial}{\partial \varphi}. \label{op:Pertthph_esc}
\end{eqnarray}

The form of the angular operator is very suggestive to expand the function in terms of the spherical
harmonics,  
\begin{equation}
\phi=\sum\limits_{lm}\,E_{l,m}(t,r)\,{Y_{0}}^{l,m}(\theta, \varphi),
\end{equation}
and as, $\square_{\theta\varphi}\,{Y_{0}}^{l,m}=\left(-l\,\left(l+1\right) +
2\,i\frac{m\,a}{r}\right)\,{Y_{0}}^{l,m}$, and $2\,a\,\frac{\partial^2}{\partial r \partial
\varphi}\,{Y_{0}}^{l,m}={Y_{0}}^{l,m}\,2\,i\,a\,m\frac{\partial}{\partial r}$. We get again an
almost radial temporal equation with only an angular coefficient  $a^2\,\cos^2\theta$ of the second
temporal derivative. 

Using the orthonormalization properties of the spherical harmonics and the fact that
$\oint\,{Y_{0}}^{l,m}\,{Y_{0}}^{l^\prime,m^\prime}\,{Y_{0}}^{2,0}=\frac12\,\sqrt{\frac{5}{\pi}}
\sqrt{\left(2\,l+1\right)\,\left(2\,l^\prime+1\right)}\,\left(\begin{array}{l l l}
l& l^\prime &2\\0&0&0\end{array}\right)\,\left(\begin{array}{l l l}
l& l^\prime &2\\m&m^\prime&0\end{array}\right)$, where we are using the Wigner-3 matrices
\cite{Miguel3p1}, again we obtain that we can completely get rid of the angular dependence in favor
of a system with coupled modes:
\begin{equation}
 \square\phi=-a^2\,As_{l+2,m}\frac{\partial^2}{\partial
t^2}\,E_{l+2,m} + \square_s\,\,E_{l,m} -a^2\,As_{l,m}\frac{\partial^2}{\partial
t^2}\,E_{l-2,m}
\end{equation}
where we have now defined
\begin{eqnarray}
As_{l,m}&=&\frac{1}{2l-1}\sqrt{\frac{(l+m)(l+m-1)(l-m)(l-m-1)}{
(2l+1)(2l-3)}} \\
{\square}_s&=&-\left(r^2  +2\,M\,r + \frac{a^2}3\,
\left(1 + 2\,\frac{l^2 + l - 3\,m^{2}}{(2l+3)\,(2l-1)}\right)\right)\,\frac{
\partial^2}{\partial t^2} +  \Delta\,
\frac{\partial^2}{\partial r^2} + 4\,M\,r\,\frac{\partial^2}{\partial t \partial r}  \nonumber  \\
&&  -2\,M\,\frac{\partial}{\partial t} + 
2\,\left(M - \frac{a^2}{r} + i\,a\,m \right)\,\frac{\partial}{\partial r}
+ 2\frac{M\,r - a^2 - i\,m\,a\,r}{r^2} - l\,\left(l+1\right).\nonumber
\end{eqnarray}

In this way, we proved that the Klein-Gordon equation for a massless scalar field in a Kerr
background can be described, as the gravitational perturbation, by a radial-temporal system of
equations for the coupled modes. This system of equations has similar properties to the ones
mentioned with respect to the gravitational perturbation, and is remarkable that for the scalar
case the coupling of the modes involve only two neighbor modes, the second neighbor above and
below a given l mode number.

The description of the evolution of the scalar field presented here, clarifies some
aspects regarding the coupling among modes, which have been discussed  in the literature, see for
instance \cite{Burko09}, and will be useful to study the properties of the scalar field evolving in
this background, such as the late time behavior known as tails. 

\section{Electromagnetic field}

The description of the electromagnetic field in a Kerr background, can also be described
within the
present formulation. Starting from the Maxwell's equations,
${F^{\mu\nu}}_{;\nu}=4\,\pi\,J^\mu$,
and considering that the Faraday tensor can be expressed in terms of the tetrad vectors
as
\cite{Chandra83}:
\begin{equation}
F^{\mu\nu}=2\,\left(\phi_0\,{m^*}^{[\mu}\,k^{\nu]} + \phi_2\,l^{[\mu}\,m^{\nu]} +
\phi_1\,\left(k^{[\mu}\,l^{\nu]}+ m^{[\mu}\,{m^*}^{\nu]}\right)\right) + {\rm c. c.}, 
\end{equation}
where $\phi_0=F_{\mu\nu}\,l^\mu\,m^\nu, \phi_2=F_{\mu\nu}\,{m^*}^\mu\,k^\nu$ and
$\phi_1=\frac12\,F_{\mu\nu}\,\left(l^\mu\,k^\nu + {m^*}^\mu\,m^\nu\right)$ are complex
scalars.
Projecting the Maxwell's equations onto a null tetrad, we obtain for
type D spaces, 
the following set of equations for the complex scalars:
\begin{eqnarray}
\left(\eta\,{\bf D} - 2\,\rho\right)\,\phi_1 - \left(\eta\,{\bf \delta}^* -
2\,\alpha + \pi \right)\,\phi_0&=&4\,\pi\,J_l, \\
\left(\eta\,{\bf \delta} + 2\,\beta - \tau\right)\,\phi_2 - \left(\eta\,{\bf \Delta} +
2\,\mu \right)\,\phi_1&=&4\,\pi\,J_k, \\
\left(\eta\,{\bf \delta} - 2\,\tau\right)\,\phi_1 - \left(\eta\,{\bf \Delta} +
\mu - 2\,\gamma \right)\,\phi_0&=&4\,\pi\,J_m, \\
\left(\eta\,{\bf D} - \rho + 2\,\epsilon \right)\,\phi_2 - \left(\eta\,{\bf \delta}^* +
2\,\pi \right)\,\phi_1&=&4\,\pi\,J_{m^*},
\end{eqnarray}
where $J_a$ are the projections of the current vector, $J_\mu$ on the respective null
vector, and $\eta=\pm 1$ depending on the choice of signature (for the present work,
$\eta=-1$).

Following the usual procedure, \cite{Teu73}, of acting with specific operators on the
equations
and using the commutation relations, we get the following equations for $\phi_0$ and
$\phi_2$:
\begin{eqnarray}
\left[\left(\eta\,{\bf D} - 2\,\rho - \rho^* - \epsilon +
\epsilon^*\right)\,\left(\eta\,{\bf
\Delta} + \mu - 2\,\gamma\right) -
\left(\eta\,{\bf \delta} - \alpha^* - \beta + \pi^* - 2\,\tau
\right)\,\left(\eta\,{\bf \delta}^* -
2\,\alpha + \pi \right)\right]\,\phi_0&=&4\,\pi\,J_0, \\
\left[\left(\eta\,{\bf
\Delta} + 2\,\mu + \mu^* + \gamma - \gamma^*\right)\,\left(\eta\,{\bf D} - \rho +
2\,\epsilon \right) -
\left(\eta\,{\bf \delta}^* +
\alpha + \beta^* + 2\,\pi - \tau^*\right)\,\left(\eta\,{\bf \delta} + 2\,\beta -
\tau
\right)\right]\,\phi_2&=&4\,\pi\,J_2,
\end{eqnarray}
and we have defined
\begin{eqnarray}
J_0&=&\left(\eta\,{\bf \delta} - \alpha^* - \beta + \pi^* - 2\,\tau
\right)\,J_l - \left(\eta\,{\bf D} - 2\,\rho - \rho^* - \epsilon +
\epsilon^*\right)\,J_m, \\
J_2&=&\left(\eta\,{\bf
\Delta} + 2\,\mu + \mu^* + \gamma - \gamma^*\right)\,J_{m^*} -
\left(\eta\,{\bf \delta}^* +
\alpha + \beta^* + 2\,\pi - \tau^*\right)\,J_k.
\end{eqnarray}

A direct substitution of the operators and spinor coefficients for the Kerr background
described by
the line element given in Eq.~(\ref{eq:lel_kerr}) using the null tetrad given by
Eq.~(\ref{tet_kerr}), after simplification produces the following equations:
\begin{eqnarray}
\left[{}_{\pm 1}\square 
+ {}_{\pm 1}\square_{\theta\,\varphi} \right]\,{\hat \phi}_{\pm 1}=4\,\pi\,J_{\pm 1}.
\label{eq:EmKS}
\end{eqnarray}
where we have defined ${\hat \phi}_1=\left(r-i\,a\,\cos(\theta)\right)^2\,\phi_0,
{\hat \phi}_{-1}=r\,\phi_2,
J_1=-2\,\Sigma\,\left(r-i\,a\,\cos(\theta)\right)^2\,J_0, J_{-1}=-2\,r\,\Sigma\,J_2$, and
the
operators:
\begin{eqnarray}
{}_1\square&=&-\left(r^2 + a^2\,\cos^2\theta +2\,M\,r\right)\,\frac{\partial^2}{\partial
t^2} + 
\Delta\,
\frac{\partial^2}{\partial r^2} + 4\,M\,r\,\frac{\partial^2}{\partial t \partial r}
 + 2\,a\,\frac{\partial^2}{\partial r \partial \varphi} -
2\,\left(r + i\,a\,\cos\theta\right)\,\frac{\partial}{\partial t}, \\
{}_{-1}\square&=&-\left(r^2 + a^2\,\cos^2\theta
+2\,M\,r\right)\,\frac{\partial^2}{\partial t^2} + 
\Delta\,
\frac{\partial^2}{\partial r^2} + 4\,M\,r\,\frac{\partial^2}{\partial t \partial r}
 + 2\,a\,\frac{\partial^2}{\partial r \partial \varphi} +
2\,\left(r + i\,a\,\cos\theta\right)\,\frac{\partial}{\partial t} + \nonumber \\
&&2\,\left(r - \frac{a^2}{r}\right)\,\frac{\partial}{\partial r} - 2\,
\frac{a}{r}\,\frac{\partial}{\partial \varphi} + 2\,\frac{a^2}{r^2},\\
{}_{\pm 1}\square_{\theta\varphi}&=& \frac{\partial^2}{\partial \theta^2} 
+ \frac1{\sin^2\theta}\,\frac{\partial^2}{\partial \varphi^2} +
\cot\theta\,\frac{\partial}{\partial \theta} \pm 
2\,i\frac{\cos\,\theta}{\sin^2\theta}\,\frac{\partial}{\partial \varphi} -
\frac1{\sin^2\theta}.
\label{op:Pertthph_em}
\end{eqnarray}

Thanks to the tetrad we are using, Eq.~(\ref{tet_kerr}), we again obtain that the main
angular
operators, ${}_{\pm 1}\square_{\theta\varphi}$, are such that the spin weighted spherical
harmonics, ${Y_{\pm 1}}^{l,m}$, are their eigenfunctions. Indeed, ${}_{\pm
1}\square_{\theta\varphi}\,{Y_{\pm 1}}^{l,m}=-l\,\left(l+1\right)\,{Y_{\pm 1}}^{l,m}$. As
$\frac{\partial}{\partial \varphi}\,\,{Y_{\pm 1}}^{l,m}=i\,m\,\,{Y_{\pm 1}}^{l,m}$, we
expand
the electromagnetic scalars in terms of such harmonics:
\begin{equation}
{\hat \phi}_{\pm 1}=\sum\limits_{lm}\,{}_{\pm 1}{\cal R}_{l,m}(t,r)\,{Y_{\pm
1}}^{l,m}(\theta,
\varphi).
\end{equation}
Substituting in the field equations, Eqs.(\ref{eq:EmKS}) we get, as for the other fields,
equations
where the angular dependence is reduced to the coefficient $\cos^2\theta$ for 
$\frac{\partial^2}{\partial t^2}\,{}_{\pm 1}{\cal R}_{l,m}$, and the coefficient
$\cos\theta$ for
$\frac{\partial}{\partial t}\,{}_{\pm 1}{\cal R}_{l,m}$. We expand these trigonometric
functions as
spherical harmonics and, using the orthonormal properties with the Wigner matrices, we
are able to
express this final angular dependence in favor of a coupling of the modes, obtaining the
following
system of equations:
\begin{eqnarray}
&-2\,a^2\,\tilde{A}_{l+2,m}\,\partial_{tt}\,{}_{\mp 1}{\cal R}_{l+2,m} \mp
4\,a\,\tilde{B}_{l+1,m}\,
\left(a\,\frac{m}{\sqrt{\left(l+2\right)\,l}}\partial_{tt} -
2\,i\,\sqrt{\left(l+2\right)\,l}\partial_{t}\right)\,{}_{\mp 1}{\cal
R}_{l+1,m} + {}_{\mp 1}\overline{\square}_{l,m}\,{}_{\mp 1}{\cal R}_{l,m} &  
\label{eq1:ev_cou_mod_em} \\
&\mp
4\,a\,\tilde{B}_{l,m}\,\left(a\,\frac{m}{\sqrt{\left(l+1\right)\,\left(l-1\right)}}
\partial_{tt} - 
2\,i\sqrt{\left(l+1\right)\,\left(l-1\right)}\partial_{t}\right)\,{}_{\mp 1}{\cal
R}_{l-1,m}
-
2\,a^{2}\,\tilde{A}_{l,m}\,\partial_{tt}\,{}_{\mp 1}{\cal R}_{l-2,m}=0,&\nonumber
\end{eqnarray}
where we have now defined
\begin{eqnarray}
\tilde{A}_{l,m}&=&\frac{1}{(2l-1)}\sqrt{\frac{(l+m)(l+m-1)(l-m)(l-m-1)(l+1)(l-2)
} { (2l+1)\,l\,(l-1)\,(2l-3)}},\\
\tilde{B}_{l,m}&=&\frac{m}{l}\,\sqrt{\frac{(l+m)(l-m)}{(2l+1)(2l-1)}},\\
{}_{1}\overline{\square}_{l,m}&=&-\left(r^2  +2\,M\,r + \frac{a^2}3\,
\left(1 +
4\,\frac{(l\,(l+1)-3)\,(l\,(l+1)-3m^{2})}{(2l+3)\,(2l-1)}\right)\right)\,\frac{
\partial^2}{\partial t^2} +  \Delta\,
\frac{\partial^2}{\partial r^2} + 4\,M\,r\,\frac{\partial^2}{\partial t \partial r}  -
\nonumber  \\
&&  2\,\left(r - 4\,i\,a\,\frac{m}{l(l+1)}\right)\,\frac{\partial}{\partial t} + 
2\,i\,a\,m\,\frac{\partial}{\partial r} - l\,\left(l+1\right),
\nonumber \\
{}_{-1}\overline{\square}_{l,m}&=&-\left(r^2  +2\,M\,r + \frac{a^2}3\,
\left(1 +
4\,\frac{(l\,(l+1)-3)\,(l\,(l+1)-3m^{2})}{(2l+3)\,(2l-1)}\right)\right)\,\frac{
\partial^2}{\partial t^2} +  \Delta\,
\frac{\partial^2}{\partial r^2} + 4\,M\,r\,\frac{\partial^2}{\partial t \partial r}  +  
\nonumber
\\
&&  2\,\left(r + 4\,i\,a\,\frac{m}{l(l+1)}\right)\,\frac{\partial}{\partial t} + 
2\,\left(r - \frac{a^2}{r} + i\,a\,m\right)\,\frac{\partial}{\partial r} -
2\,i\,m\frac{a}{r} +
2\frac{a^2}{r^2} - l\,\left(l+1\right).
\end{eqnarray}

Thus, the dynamics of the electromagnetic field in a Kerr
background can be described, as the other fields, by a radial-temporal
system of
equations for the coupled modes. This system of equations has similar properties to the
ones
mentioned with respect to the gravitational perturbation.

\bibliographystyle{unsrt}
\bibliography{refs}

\end{document}